# Population-balance description of shear-induced clustering, gelation and suspension viscosity in sheared DLVO colloids

Marco Lattuada[&], Alessio Zaccone[$], Hua Wu[§] and Massimo Morbidelli[§]

[&]Adolphe Merkle Institute, University of Fribourg, Chemin des Verdiers 4, 1700 Fribourg, Switzerland

[$]Department of Chemical Engineering and Biotechnology, Cambridge University, Pembroke Street, Cambridge CB2 3RA, UK

[§]Institute for Chemical and Bioengineering, Department of Chemistry and Applied Biosciences, ETH Zurich, Vladimir-Prelog-Weg 1, 8093 Zurich, Switzerland

Corresponding authors:

marco.lattuada@unifr.ch

massimo.morbidelli@chem.ethz.ch

az302@cam.ac.uk

## Abstract

Application of shear flow to charge-stabilized aqueous colloidal suspensions is ubiquitous in industrial applications and as a means to achieve controlled field-induced assembly of nanoparticles. Yet, applying shear flow to a charge-stabilized colloidal suspension, which is initially monodisperse and in quasi-equilibrium leads to non-trivial clustering phenomena (and sometimes to a gelation transition), dominated by the complex interplay between DLVO interactions and shear flow. The quantitative understanding of these strongly *nonequilibrium* phenomena is still far from being complete. By taking advantage of a recent shear-induced aggregation rate theory developed in our group, we present here a systematic numerical study, based on the governing master kinetic equation (population-balance) for the shear-induced clustering and breakup of colloids exposed to shear flow. In the presence of sufficiently stable particles, the clustering kinetics is characterized by an initial very slow growth, controlled by repulsion. During this regime, particles are slowly aggregating to form clusters, the reactivity of which increases along with their size growth. When their size reaches a critical threshold, a very rapid, explosive-like growth follows, where shear forces are able to overcome the energy barrier between particles. This stage terminates when a dynamic balance between shear-induced aggregation and cluster breakage is reached. It is also observed that these systems are characterized by a cluster mass distribution that for a long time presents a well-defined bimodality. The model predictions are quantitatively in excellent agreement with available experimental data, showing how the theoretical picture is able to quantitatively account for the underlying *nonequilibrum* physics.

# Introduction

Aggregation of colloidal suspensions is a physical phenomenon with a widespread range of applications in nanotechnology, materials science, food science, biomedical science and waste-water treatment [1]. While for some systems aggregation is highly deleterious and must be avoided at all costs, in self-assembling nanoparticles it is a crucial step[2-4]. It is therefore not surprising that in order to better design colloidal dispersions, the fundamental understanding of aggregation phenomena has been and still is the subject of many investigations, both experimental and theoretical[5-10]. The number of studies on aggregation has increased exponentially in the last three decades. In spite of the enormous progress made to fill the knowledge gap, some phenomena are still poorly understood, in particular a bottom-up approach connecting the physics and chemistry of colloidal particles with the time-dependent clustering behavior and the macroscopic rheology is currently missing [11].

Two situations have been systematically investigated. Quiescent colloidal dispersions have been the subject of the majority of experimental and theoretical work, culminated with the discovery of two universal aggregation regimes: diffusion-limited aggregation, where particles only experience short range attractive interactions, and reaction-limited aggregation, where particles also experience repulsive interactions, typically of electrostatic origin [5]. These two regimes differ substantially not only in terms of kinetics, but also of aggregate structures[6, 12]. The second system that has been often considered, for its practical relevance, is aggregation of colloidal dispersions in the presence of shear forces[8, 9, 13-28]. Shear-flow has a strong accelerating effect on the aggregation kinetics, and a considerable influence on the structure of the clusters formed, but it also affects cluster size by breaking them up [8].

Shear-induced aggregation has been investigated in the past with fully destabilized suspensions[14, 15, 17-25, 27, 28]. A much more complex and physically rich scenario, with enormous practical implications, is the behavior of colloidal suspensions subject to shear forces and simultaneously stabilized by repulsive electrostatic interactions[7, 9, 13, 16, 26, 27, 29-32]. These systems have competing interactions (attractive van der Waals at short-range, repulsive double-layer at larger separation), a situation often encountered also in protein systems[33-35]. In the absence of external fields, and at sufficiently dilute conditions, strongly charge-stabilized systems are well described by equilibrium statistical mechanics. If the electric double-layer repulsion is substantial, aggregation is extremely slow, and for quite a long time the radial distribution function goes approximately as predicted by equilibrium statistical mechanics, $g(r) = \exp[-U/kT]$, where $U$ is the repulsive part of the DLVO pair-potential. Application of shear flow allows the particles to explore the van der Waals attractive minima of the DLVO interaction, and perturbs this initial quasi-equilibrium state. Detailed-balance in the collisions between particles is broken (due to most flow-induced collisions being de facto irreversible, for which the reverse rate, breakup, is practically zero), and by constantly injecting energy into the system, thus bringing it far away from thermodynamic equilibrium [36]. It is a fundamental question of statistical physics to elucidate the time-evolution of such a strongly driven system, and to determine which nonequilibrium steady-states act as attractors for the dynamics at long times. Hence, making predictions about the time-evolution of such strongly nonequilibrium systems represents a major challenge in modern statistical physics[37].

At the colloid-particle level, these systems have a unique behavior, since the presence of repulsive interactions stabilizes particles against shear forces as long as the latter are not providing sufficient energy to overcome the repulsive barrier[30-32]. In this event, the aggregation rate becomes very high, reaching levels comparable to those observed for particles in the absence of repulsive interactions. This peculiar effect, which has been addressed both theoretically[30, 31] and experimentally[32], manifests itself with an explosive-like, runaway behavior in the growth rate of the cluster size. Initially, the system appears to undergo an almost negligible aggregation, which is however followed by a regime where an exponentially fast, auto-catalytic cluster growth is observed[30].

The objective of this work is to present a bottom-up quantitative description of the time-dependent evolution of DLVO colloids in shear flow, starting from an initial equilibrium state (stable colloidal suspension) and predicting the *nonequilibrium* cluster size evolution under account of both shear-induced cluster aggregation and breakup. This task is achieved by numerically solving the governing master equation (population balance) with physically justified microscopic kernels for aggregation and breakup. The simulations shed light on the resulting cluster mass distribution, focusing in particular on the development of a marked *bimodality*, confirmed by experiments. Such bimodality has important implications, because it implies that only two distinct populations of clusters can survive at steady-state: primary particles with very small clusters, on one end of the spectrum, and very large clusters, at the other end, whose size is determined by breakage. Finally, the obtained cluster size distribution as a function of time can be used to estimate the time-evolution of the steady shear viscosity of the suspension, for the first time in quantitative agreement with experiments, and to predict the occurrence of gelation at long times. Gelation is a possible outcome provided that the initial colloid concentration is such that the final fractal-cluster volume fraction reaches close packing.

**Simulation methodology**

All simulations performed in this work have been carried out by solving the governing master equation which we will refer to as population balance equations (PBEs), in the following discrete form [38, 39]:

$$\frac{dN_m}{dt} = \frac{1}{2}\sum_{i,j=1}^{i+j=m} K_{ij}^A N_i N_j - N_m \sum_{i=1}^{\infty} K_{im}^A N_i - K_m^B N_m + \sum_{i=m+1}^{\infty} K_i^B G_{im} N_i . \tag{1}$$

In Equation (1), $N_m$ is the number concentration of clusters with mass $m$, $K_{ij}^A$ is the aggregation rate or aggregation kernel between two clusters with masses $i$ and $j$, respectively; $K_i^B$ is the breakage rate o of a cluster with mass $i$ and $G_{im}$ is a fraction of fragments with mass $m$ produced by a breakage of a cluster with mass $i$, with $i \geq m+1$. All information about the physics of the aggregation and breakage processes are contained in the kernels. The following expression, based on the analytical solution to the Smoluchowski (advection-diffusion) equation with shear for arbitrarily interacting Brownian particles developed in Refs.[30-32], has been used to model the rate of shear-induced aggregation between two clusters with masses $i$ and $j$

$$K_{ij} = \min \begin{cases} \frac{2kT}{3\eta W}\left(i^{\frac{1}{d_f}} + j^{\frac{1}{d_f}}\right)\left(i^{-\frac{1}{d_f}} + j^{-\frac{1}{d_f}}\right)e^{2\alpha Pe} \\ a\frac{4}{3}\dot{\gamma}R_p^3\left(i^{\frac{1}{d_f}} + j^{\frac{1}{d_f}}\right)^3 \end{cases} \tag{2}$$

where $W$ is the colloidal stability ratio, $\dot{\gamma}$ is the shear rate, $\eta$ the viscosity of water, $k$ the Boltzmann constant, $T$ the temperature, $d_f$ is the cluster fractal dimension, $R_p$ the primary particle radius, $i$ the mass of the $i^{th}$ cluster, $\alpha=1/3/\pi$ and $a$ is the collision efficiency. The Fuchs stability ratio is given by the following equation [1]:

$$W = 2\int_{2}^{\infty} \frac{\exp\left(\frac{U(l)}{kT}\right)}{G(l)\cdot l^2} \mathrm{d}l \qquad (3)$$

where $U(l)$ is the interaction energy between a pair of particles located at a dimensionless distance $l$ (i.e., the particle center-to-center distance normalized by the primary particle radius) and $G(l)$ is a hydrodynamic resistance function accounting for the reduction in diffusivity when two particles are moving towards one another. Quite often, the Fuchs stability ratio is simplified as follows:

$$W \simeq \frac{1}{2\kappa R_p} \exp\left(\frac{U_{Max}}{kT}\right) \qquad (4)$$

where $U_{Max}$ is the interaction energy barrier height and $\kappa$ the inverse Debye length. The collision efficiency has been evaluated according to the model developed by Bäbler[40]. The functional form of the kernel is basically identical to the one proposed in[30] and[32].

The Peclet number is defined as follows:

$$Pe = \frac{3\pi\eta\dot{\gamma}R_{H,i}R_{H,j}\left(R_{H,i}+R_{H,j}\right)}{2kT}, \qquad (5)$$

where $R_{H,i}$ is the hydrodynamic radius of the $i^{th}$ cluster, which has been computed using correlations developed in our group [41]. The minimum in Equation (2) signifies that the kernel is equal to the traditional shear kernel when the first part containing the exponential term in Equation (2) becomes higher than the shear kernel. The viscosity used in Equation (5) is the viscosity of water. The peculiar form of the aggregation kernel, Equation (2), comes from the thermally-activated Arrhenius-like competition between shear forces and repulsive interactions (stemming from the analytical solution to the Smoluchowski equation with shear [26]), which gives rise to the exponential term. The physical consequence of the initial exponential dependence of the aggregation rate on $Pe$ in Equation (2) is a very strong sensitivity of the aggregation kernel on the cluster size. Depending on the strength of electrostatic repulsion among particles, which is contained in the Fuchs stability ratio $W$ (see for example Equation (4)), the rate of aggregation between small clusters might be insensitive to shear and dominated by repulsion as in a reaction-limited aggregation process. On the contrary, the aggregation of large clusters is insensitive to repulsive interaction and is purely controlled by shear [26]. The exponential transition between the two regimes is the key for the interpretation of the experimental data and for the peculiar form of the cluster mass distribution discussed in detail in the following.

The rate of breakage of clusters has been modelled using a power-law model proposed in Ref. [28]:

$$K^B_i = c_1 (\eta \dot{\gamma})^n R^m_{g,i}. \qquad (6)$$

In Equation (6) $R_{g,i}$ is the radius of gyration of a cluster with mass $i$, while $c_1$, $n$ and $m$ are parameters depending on the flow field, on the primary particle size and above all on the cluster fractal dimension. The values of these parameters for the general case are reported in the literature [28]. However, the value of the prefactor $c_1$ in this work for the cluster fractal dimension value equal to *2.7* has been set to $2.38 \cdot 10^{-10}$. The fragment mass distribution has been assumed to be binary and symmetric. In all simulations, it has been assumed that the breakage mechanism is only active for clusters with a mass larger than 1000 particles, while the breakup rate is zero for all clusters with mass smaller than 1000. This condition, which clearly breaks detailed balance in the master equation Equation (1), is consistent with the observation made several times in the literature that clusters below a critical mass are not subject to breakage, especially when considering the narrow range of shear rates analyzed in this work [28, 42]. The power-law dependence of the breakup rate on the fractal cluster size, with an exponent which is a function of the fractal dimension, can be analytically justified with the framework of Conchuir and Zaccone by solving the Kramers escape rate problem for the breakup of inner bonds inside the aggregate under the action of shear [42]. The competition between shear force and colloidal binding force in the thermally-activated Kramers rate gives rise to a criterion to establish that breakup becomes a fast process when the shear energy exactly balances the binding energy. Since the shear energy depends on the cluster size with a power-law which is a function of $d_f$ and of the stress-transmission through the cluster, the criterion justifies Equation (4).

The solution of PBEs (1) has been carried out by means of the Kumar-Ramkrishna method, which allows one to cover a broad range of cluster masses [38]. Three hundred pivots have been used to cover an interval of cluster masses going from one to $10^{10}$ particles. Unfortunately, the form of the aggregation kernel (2), combined with a breakage mechanism, leads to an extremely stiff system of ordinary differential equations, the stiffness of which increases with increasing the Fuchs stability value.

The viscosity of the suspension undergoing aggregation has been simulated by using the equation proposed by Van de Ven and Takamura, which has the following form [43]:

$$\eta_s = \eta \left( \frac{1 - \frac{\phi}{\phi_c}}{1 - (k_0 \phi_c - 1)\frac{\phi}{\phi_c}} \right)^{-\frac{5\phi_c}{2(2 - k_0 \phi_c)}} \quad (7)$$

where $k_0$ is a parameter accounting for second order hydrodynamic interaction between particles, as well as for secondary electro-viscous effect and for shear thinning behavior. The expression for $k_0$ used in this work, which is rather involved, is reported in the original publication [43]. The critical volume fraction is set to $\phi_c = 0.59$. The volume fraction used is the one occupied by the clusters, which is a function of time:

$$\phi(t) = \frac{4}{3}\pi \sum_{i=1}^{\infty} N_i(t) \cdot R_{g,i}^3 \quad (8)$$

In all simulations carried out in this work, unless stated otherwise, the initial condition is an equilibrium population of monodisperse spherical particles.

## Experimental data

The experiments used to obtain the data discussed in this work have already been partially presented in a previous publication[32]. Briefly, the colloidal system used is a surfactant-free colloidal dispersion of styrene-acrylate copolymer particles in water, supplied by BASF AG (Ludwigshafen, Germany) and produced by emulsion polymerization. The particles are nearly monodisperse with mean radius, a=60±1 nm, as measured by both dynamic light scattering (Nano-ZS Malvern, UK) and small-angle light scattering (Mastersizer 2000 instrument, Malvern, UK). The particles have been cleaned by mixing with ion-exchange resins, and the surface tension of the suspension has been measured by means of the Wilhelmy plate method with a DCAT-21 tensiometer (Dataphysics, Germany). Only suspensions with surface tension ≥71.7 mN/m have been used in the experiments. For the shearing experiments, a small amount of electrolyte NaCl (17 mM) was added to make up the ionic background. This ionic strength is much lower than the critical coagulation concentration (50 mM). The electrolyte has been added in such a way as to avoid uncontrolled aggregation, and to make sure that the particles would always be in contact with electrolyte solutions at a concentration substantially smaller than the critical coagulation concentration. The original polystyrene particles suspension, which has a much higher particle volume fraction than the one used in the experiments, has been mixed with a pre-dilute solution of NaCl in MilliQ water (Merck Millipore, DE), so as to reach the desired volume fraction and ionic strength.

A strain-controlled rheometer (Rheometric Scientific) in Couette mode was used to shear the samples. The gap between the outer cylinder and the inner one is 1 mm and the length of the latter is 34 mm. The outer cylinder is temperature controlled at *T=298±0.1* K and a solvent trap has been fixed on the outer rotating cylinder to limit evaporation. The latex suspensions and NaCl solutions have been properly mixed so as to avoid heterogeneities in the concentration, which would cause irreproducible aggregation phenomena. The shearing was switched on exactly 7 min from the time of mixing between latex and background NaCl solution.

In order to confirm experimentally the bimodality of the cluster mass distribution, samples were taken from the suspension subject to stirring at defined time points, and filtered by means of a 5μm cut-off filter, in order to remove the larger cluster and determine the fraction of particles and small clusters in the system, thus permitting the determination of the conversion to large clusters.

## Results and discussion

We will start by discussing the results of the calculations obtained by solving Equations (1) in combinations with the aggregation and breakage kernels (2) and (6), respectively. In Figure S1 we show the dependence of aggregation kernel (2) on the *Pe* number, in the case of aggregation between equal-sized clusters, for three different values of the Fuchs stability ratio *W*, reported in the legend. One should note that the *Pe* number is a function of the cluster size, as Equation (5) indicates. It can be observed how the aggregation rate is almost cluster-mass independent for sufficiently low *Pe* values. This regime is then followed by steep exponential increase of the aggregation rate around a critical *Pe* value, which increases with *W*, before finally growing linearly with *Pe*, indicating a shear-controlled aggregation, completely independent of *W*. This peculiar trend has one clear consequence. If, during the aggregation process, one starts with particles that are sufficiently electrostatically-stable under a given shear rate, the initial stages of aggregation will be rather slow. However, the formation of clusters with larger size (and corresponding larger *Pe*) will lead to a progressive increase in the aggregation rate, until the critical *Pe* value is reached. At this time point, the aggregation will suddenly speed up, with an "explosive" behavior, due to the transition to auto-catalytic shear-controlled regime. However, the breakage rate will oppose the effect of aggregation more and more strongly for larger clusters, and will lead to a stable size.

*Cluster size evolution*

In order to better analyze this qualitative picture, in Figure 1 the dimensionless time evolution of the normalized average cluster radius of gyration is shown for four different values of *W*, from $10^3$ to $10^6$. The dimensionless time $\tau$ is defined as follows:

$$\tau = \frac{8kTN_0}{3\eta W} t \qquad (9)$$

where $N_0$ is the initial number concentration of primary particles and *t* is the physical time. The dimensionless time defined in Equation (9) can be used to collapse all experimental and simulated data of aggregation processes obtained under stagnant conditions on a single mastercurve[6]. In fact, the quantity used to make the physical time dimensionless in Equation (9) is the initial aggregation rate of particles, which provides the correct time scale to describe aggregation under stagnant conditions.

The calculations are carried out both with (continuous lines) and without (dashed lines) breakage. One can observe that the behavior qualitatively discussed above on the basis of the collision physics, is well reflected in the PBE calculations. The average cluster size initially grows very slowly, and the slow growth is then followed by an explosive growth, which continues until the entire mass of the system accumulates in the last bin of the cluster mass distribution in the absence of breakage. The latter regime is the shear controlled aggregation. In the presence of breakage, instead, the size reaches a plateau after the explosive growth, which is due to the dynamic balance between aggregation and breakage. By increasing the value of the Fuchs stability ratio, the explosive regime is shifted to higher physical times $t$. Nevertheless, when plotted against the dimensionless time defined by Equation (9), the various curves do not overlap, as it would happen in the case of stagnant aggregation, because the dimensionless time values where the explosive growth takes place decrease as $W$ increases. This indicates that the initial aggregation rate is not the correct time scale to describe the entire aggregation process. Additionally, the explosive growth rate becomes steeper as $W$ increases, while the plateau reached in the presence of breakage is unaffected by the value of $W$. All these information suggest that, as the height of the DLVO repulsive energy barrier increases, the crossover from slow to fast aggregation kinetics becomes sharper and more abrupt.

*Cluster mass distribution evolution*

In order to explain the behavior of the time evolution of the average cluster size, it is particularly informative to look at the time evolution of the cluster mass distributions. The results are showcased in Figures 2a and 2b for $W=10^5$. The dimensionless time values have been chosen such as to show the CMDs before, during and after the explosive growth. In the absence of breakage (Figure 2a), the cluster mass distribution becomes progressively broader as larger clusters are formed, while primary particles are depleted. The shape of the CMD is drastically altered by the presence of breakage (Figure 2b). In the slow, pre-explosive growth phase we have a similar time evolution of the cluster mass distribution, since breakage is not very active. However, during the explosive growth, when larger clusters begin to form, the breakage process stops their growth, and the cluster mass distribution develops a peak, corresponding to the average size of the clusters for which the dynamic balance between aggregation and breakage is reached. This means that, for some time, the aggregation process leads to a bimodal cluster mass distribution: on one end of the size spectrum there are primary particles and very small clusters, only. In the intermediate cluster mass range, the concentration of clusters present is negligibly low, because their aggregation rate is fast enough to be rapidly consumed by the aggregation, while the breakage process of larger clusters is not fast enough to lead to their accumulation. Instead, the largest clusters can accumulate, because beyond a critical size the rate of aggregation can be effectively counterbalanced by the breakage process. After a sufficient amount of time, the small clusters are completely consumed, so that the cluster mass distribution becomes monomodal again, with only the large clusters peak surviving. This implies that the competition of shear forces and electrostatic repulsions can lead to the formation of very large clusters, which coexist with small ones, without clusters in the intermediate size range. Two additional sets of CMDs are shown, for stability ratio values $W=10^4$ and $W=10^6$, in Figures 3a and 3b, respectively. The trend is similar to the one shown in Figure 2b and even more pronounced in the case of $W=10^6$, while for $W=10^4$ the bimodality develops later on, and is less pronounced. This shows that, whenever the stability ratio is small, the explosive growth is less prominent and the cluster mass distribution first develops a broad monomodal shape, followed only later on by the consumption of the intermediate-size clusters. All these observations provide a hint on the reason why, as shown in Figure 1, the dimensionless explosion times decrease as the stability ratio increases. High stability ratios promote the rapid development of a bimodal distribution, which induces the explosive growth, because there is a large difference between the rates of aggregation of primary particles and of large clusters, which is independent of the stability ratio.

*Effect of breakage rate and fragment mass distribution*

The solution of population balance equations with the combination of aggregation kernel (2) and breakage kernel (6) represents a significant numerical challenge, because the differential equations exhibit high numerical stiffness. We decided to devote some efforts in investigating the role played by the different parameters on the solution of the population balance equations. To this purpose, the effect of both the breakage rate and the form of the fragment mass distribution has been analyzed. The effect of changing the breakage rate is shown in Figure 4, where the time evolution of the average cluster radius of gyration is shown for $W=10^5$, and five different values of the breakage rate. The breakage rate has been varied over a few orders of magnitude by adjusting the value of the prefactor $c_1$ in Equation (6). The effect of increasing or decreasing the rate of breakage is quite interesting, and somehow counterintuitive. As the breakage rate increases, the plateau reached by the average size is lowered, which is consistent with the expectation of a faster breakage process shifting the balance between aggregation and breakage towards smaller sizes. However, a faster breakage also leads to a reduction in the time required to reach the explosive growth. This rather counterintuitive behavior is caused by the breakage process promoting the formation of fragments in the size range corresponding to the critical *Pe* number for the slow-to-fast kinetic crossover. The higher the concentration of such clusters, the faster and steeper will be the explosive behavior of the system. This fact, however, also affects the numerical stiffness of the problem, which increases with the breakage rate.

In Figure S2-S4, instead, a comparison between the cluster mass distributions is shown in the case of $W=10^5$, a fixed value of the breakage rate and three different fragment mass distributions: binary symmetric, binary asymmetric (erosion type, with a ratio between the masses of the two fragments equal to 1/10), and a broader fragment mass distribution, obtained by extrapolating to large clusters the results obtained from Stokesian Dynamic simulations [28]. The results show that a variation of the fragment mass distribution has a strong effect on the functional form of cluster mass distribution. Switching from symmetric breakage to asymmetric breakage causes a broadening of the peak corresponding to large clusters, which is not unexpected, because a larger number of small clusters will be produced as a result of the breakage process. When moving to an even broader fragment mass distribution, the entire shape of the cluster mass distribution is modified, and the bimodality disappears almost entirely, substituted by a continuous very broad cluster mass distribution, similar to that found in the absence of breakage. This is an important observation, implying that the shape of the CMD could provide valuable information about the details of the breakage mechanism.

Due to practical limitations, the experimental data available to test this kernel have been obtained with particles having a high colloidal stability. The values of Fuchs stability ratio *W* are close to $10^8$. This means that the PBEs with the two aforementioned kernels are too stiff to be solved numerically. Therefore, a different approach has been used. We have introduced a simplified but effective pseudo breakage mechanism, in order to compare the results of our simulations with experimental data. Since the breakage rate predicted by Equation (6) increases strongly with the cluster mass, it has been assumed that, instead of a power-law dependence over the entire cluster mass, the breakage rate is infinitely high, in practice, above a critical cluster mass. This is modelled by effectively imposing a zero rate of aggregation of clusters above the same critical cluster mass, together with a finite breakage rate of clusters, to prevent their unphysical accumulation. This allows one to simulate a fast breakage mechanism above a certain mass threshold. Since the steady state size for a given stability ratio depends on the shear, the scaling of the steady state size as a function of the applied shear rate has been determined for a few low values of the stability ratio by solving the full model. The dependence has then been extrapolated to the high stability ratio values, where the solution of the full model was impossible. From a physical point of view, setting the breakage rate to infinity above a threshold is justified: fractal clusters become less and less dense, hence less and less mechanically stable, as they grow, because the mechanical stability is controlled by the inter-particle connectivity which decreases upon decreasing the inner density of the cluster. Eventually, a maximum mechanically-stable size is reached for which the breakup rate has a vanishing activation energy and breakup is a fast process for all cluster sizes above the threshold [42].

Some tests were performed to see under which conditions the predictions of this approach could match that of the rigorous solution of the PBE, with kernels (2) and (6) applied over the entire CMD. Figure S5 shows this comparison. Figure S5a shows the time evolution of the average cluster radius of gyration, while Figure S5b shows a comparison of viscosity profiles as a function of time. One can observe that, by properly selecting the critical size above which breakage is instantaneous, the two approaches lead to similar results, in terms of average cluster size and evolution of viscosity in time. In Figures S5c and S5d it is also shown that the shape of CMDs remains qualitatively similar, even though some quantitative differences are observed. Therefore, for the comparison with experimental data, this simplified approach will be used.

*Comparison with experimental data*

*1. Viscosity and cluster radius of gyration*

The comparison of PBEs-based calculations with experimental data, some of them already published by Zaccone et al. [32], is discussed in the following. Figure 5 presents the time evolution of the viscosity profiles as a function of time, for a few conditions, with particle volume fraction values ranging from 19 to 23% and a few shear rates, as indicated in the legend. In all cases, the viscosity of the suspension remains almost constant and equal to the initial viscosity for a certain lag time, followed by a very rapid growth. Using just one single fitting parameter, *i.e.*, the Fuchs stability ratio $W$, the values of which have been judiciously fixed within the range expected from DLVO theory, to $1.38 \cdot 10^8$, $10^8$ and $6.5 \cdot 10^7$ at volume fractions of 19, 21 and 23%, respectively, all the time evolution profiles can be well predicted by the PBE calculation. The stability ratio values have been obtained by fitting the viscosity evolution profiles for the following cases: 19% particle volume fraction and a shear rate of 1700 s$^{-1}$, 21% particle volume fraction and a shear rate of 1700 s$^{-1}$ and 23% particle volume fraction and a shear rate of 1300 s$^{-1}$. The small decrease in $W$ with increasing the particle volume fraction can be justified on the basis of the colloid concentration effects on the colloidal stability of the dispersion. Given the high values of $W$ obtained from the fitting, it appears impossible to proceed with determining W independently, for example by measuring initial aggregation kinetics under stagnant conditions, since the kinetics would be too slow to be detectable.

In Figure S6 the time evolution of the occupied volume fraction by the clusters computed from Equation (8) is shown, for the same set of data shown in Figure 5. One can observe how the occupied volume fraction shows the same trend as the viscosity, and tends to reach the critical threshold of 1 for the same time value where the viscosity diverges. The value of 1 is critical for the volume fraction, because it is indicative of clusters occupying the entire volume available, thus causing percolation and gelation. One should notice that gelation is usually reached when the tail of the cluster mass distribution appears, described by a power-law, causing a divergence in the average cluster mass [44]. In the present case, instead, such tail in the cluster mass distribution is absent, and gelation is instead the result of the progressive accumulation of large clusters until random close packing is reached.

For one specific condition, i.e., 21% and a shear rate of 1700 s$^{-1}$, some data about the size evolution (average cluster radius of gyration measured by static light scattering) as a function of time are available, and shown in Figure 6. The first striking feature of these data is that the rapid growth in the average cluster size occurs at a time of about 1500 s, while the viscosity explosive increase occurs at about 7200 s. The mismatch is due to the strong sensitivity of light scattering data to the presence of clusters compared to viscosity. While a few clusters are sufficient to be detected by SLS, viscosity starts to be affected only at a much higher conversion of particles into clusters. One should note that inside the rheometer, the shear rate is never perfectly uniform. Therefore, a few large clusters could have been created in those small regions where the shear rate is higher than the average value. In addition to the experimental data, the numerical predictions of population balance equations calculations are shown in the same figure. The maximum size reached by the clusters is a quantity that depends on the dynamic balance between breakage and aggregation. With the simulation approach used for these calculations, this value has been set by the limiting cutoff value of mechanically-stable cluster mass beyond which no aggregation occurs (because of instantaneous breakage of the mechanically-unstable large aggregates). The results of the calculation indicate that the model captures only qualitatively the experimental trend. While the model captures the existence of a delay in the observed explosive growth of the viscosity compared to the size, it significantly underestimates such delay. The model predicts the explosive growth in the average cluster size at a time of about 5000 s, while the prediction of the viscosity growth time is accurate. The mismatch between model predictions and experimental data for the actual values could have several explanations. First of all, the presence of clusters generated in higher shear regions of the rheometer could explain the early explosion of the cluster size. Such clusters cannot be predicted by simulations. Additionally, simulations have been carried out with a constant cluster fractal dimension, while experimental data indicate that the first clusters have a lower fractal dimension, which increases because of shear forces quickly to the asymptotic value of 2.7 [45]. More open clusters have a higher collision radius, which could increase the overall rate of aggregation and reduce the time to explosion. However, an implementation in the population balance equations of time-dependent fractal dimension requires the knowledge of a law describing the time evolution of the cluster structure, which is somehow elusive and usually semi-empirical [45]. Such simulations would also be extremely time-consuming. Additionally, the mismatch could also be caused by overestimating the rate of aggregation of larger clusters with small particles, possibly as a result of neglecting many-body hydrodynamic interactions. Finally, the aggregation models developed so far apply to dilute conditions, while the experimental data available have been obtained at quite high volume fraction. Simulation results obtained in stagnant conditions indicate that the increase in concentration has strong effects on the aggregation mechanism [46].

*2. Conversion of primary particles to clusters*

Figure 7 shows data on the conversion of particles to clusters, together with model predictions. One can observe that the conversion of particles into clusters is relatively slow at the beginning, and undergoes a rapid acceleration approximately at the same time as the viscosity. This is consistent with the overall picture that the formation of large clusters, which are responsible for causing the explosion in the average cluster size, only involves a relatively small fraction of the total initial number of primary particles. The viscosity changes only when the conversion increases substantially. Two model predictions are shown. The first one has been obtained by considering the conversion of particles to clusters of any size, including dimers, while the second prediction is the conversion to clusters containing three particles or more. The large difference between the two predictions indicates the sensitivity of conversion to clusters, and shows how important the contribution of dimers is to the overall conversion. Since the two model predictions are bracketing the experimental data, it can be concluded that the model can capture the conversion profile well, given the difficulty of experimentally separating primary particles and oligomers from larger aggregates.

*3. Analysis of cluster mass distribution*

It is also instructive to look at the computed CMDs at different time points, in order to observe its evolution at: i) before the size explosive growth, ii) at a time point after the size explosive growth but before the viscosity explosive growth, and iii) after the viscosity explosive growth. These results are shown in Figure 8. The size explosion time is about 5000s, while the viscosity explosion time is about 7200s. One can observe that, for time values lower than the size explosion time, the CMD is monomodal, with primarily primary particles and very few small clusters. As soon as the size explosion time is reached, a second cluster population appears, with masses close to the cutoff values. From that point on, the second population of cluster grows substantially in number, consuming the small particles. For time values above the viscosity explosion time, the small particles are rapidly consumed, before disappearing completely. One should further highlight how clusters with intermediate size, comprised between the two populations, keep having exceedingly low concentrations, since their formation is rapidly compensated by their consumption to generate larger clusters, for which further growth is prevented by the presence of cluster breakage. In order to compare model predictions with experimental data, In Figure 9a and 9b the experimentally measured scattering structure factors are reported for the same experimental condition already discussed in Figures 6 and 7. Figure 9a shows the full scattering structure factors evolution as a function of time, where it appears that size of the clusters grows rapidly to a steady state, after which remains almost unchanged, while the progressive growth of the intensity demonstrates the increase in the number of large clusters. Figure 9b, instead, shows some examples of structure factors after filtration with a filter having a 5 micrometers pore size. The structure factors are relatively flat, consistent with the presence of only a tiny fraction of small clusters. The same figure shows the predictions of the model, computed by excluding all clusters with a size larger than 5 micrometers. The model predictions are consistent with the experimental data, and show that only a small fraction of small clusters are present.

Figure 9c shows on the other hand the model calculations of the average structure factor evolution as function of time, for the same conditions shown in Figure 9a. Note that the time points for which the structure factor has been calculated are not the same as the measured ones. The comparison between model predictions and experimental data for the kinetics of average cluster size shown in Figure 6 has already evidenced a mismatch between measured and predicted size growth. The calculated times have been chosen in order to show that the main features of structural evolution of the calculated structure factors are consistent with the experimental ones. In particular, it is important to highlight that the growth of the structure factor predicted by the model presents some unique features. Both Figures 9a and 9c show that the structure factor grows starting from the low $q$ range, while the intensity in the high $q$ range remains initially unaffected. Only after substantial growth of the clusters, the intensity in the high q range will begin to increase. This type of growth is completely different from the one predicted in the case of any other aggregation mechanism. As an example, we showed in Figure S7 the growth of the scattering structure factor in the case of diffusion-limited aggregation and shear-induced aggregation (with and without breakage). In all of these cases, one can observe that the time evolution of the scattering structure factor is very different from the one observed in Figure 9a and 9c. In all of the cases shown in Figure S7 the structure factor grows uniformly as a function of time over the entire $q$ range. The only growth mechanism compatible with the observed growth pattern of the scattering structure factor is the explosive kernel discussed in this work. This observation clearly supports the mechanism proposed to explain the set of data.

*4. Viscosity explosion (gelation) time at different volume fractions*

Finally, Figure 10 shows the dependence of the explosion time (determined from the viscosity profile) on the shear rate, for three experimentally measured volume fractions. The experimental data are compared to the numerical calculations. In reference [32] it was shown that the dependence of the explosion times as a function of the shear rate have primarily an exponential dependence, which is the same dependence of particles aggregation rate on the applied shear in the presence of repulsive interactions. The numerical model is able to predict quantitatively the explosion times for all conditions in a slightly more accurate fashion. The scaling is not a simple exponential dependence, because the time of explosion is not only affected by the change in shear rate, but also by the change in the maximum size reached as a result of the balance between aggregation and breakage.

It is also important to emphasize here that the time at which the viscosity explodes and seemingly diverges, coincides with the transition of the suspension from liquid-like into a solid-like gel. This is confirmed by the observation that the runaway of viscosity leads, in the experimental setup, to the arrest of the rheometer and the material has all the appearance of a soft gel with a finite shear modulus and a storage modulus much larger than the loss modulus. One should further notice that the formed gels are irreversible, and do not turn back to liquid state upon ending the application of shear. This rules out any possibility that the formed jammed state is due to hydroclusters, as shown by Stokesian Dynamic simulations of stable colloidal suspensions undergoing shear thickening.[47] The gelation transition is caused by the jamming of the clusters, which reach close-packing. The phenomenon is made possible by the fact that clusters are fractal, hence upon growing they occupy an effective packing fraction which effectively increases and may reach close-packing, as in our case, if the initial volume fraction of colloids is larger than a threshold (for the standard case of DLCA or RLCA gelation in quiescent conditions this is achieved at vanishing colloid concentration owing to the much lower fractal dimension of the clusters). Population balance calculations confirm this effect, as was mentioned in discussing Figure S6.

## Conclusions

Charge-stabilized suspensions of DLVO-interacting colloidal particles which are initially monodisperse and in a quasi-equilibrium state, can be driven into irreversible clustering upon application of an external steady shear flow. The fundamental question about the time-evolution of the shear-induced aggregation process is crucial for many applications (from directed self-assembly in nanotechnology to many industrial processes) as well as for our basic understanding of *nonequilibrium* irreversible processes. The irreversible dynamics of the process may in fact reach non-trivial *nonequilibrium* steady-states such as gelation, at which point the dynamics arrests and the viscosity diverges[48, 49].

As is well known, the main tool to quantitatively study this type of problems is offered by master kinetic equations, for which analytical solutions are known only for very few special cases, while numerical solutions may often be also challenging due to nonlinearities and numerical stiffness. Here we presented a numerical solution to the master equation (population balance equation) governing the shear-induced aggregation of DLVO colloids at steady-shear, using a physically justified aggregation rate theory previously formulated by our group that can capture the essential physics of the process [30]. From the numerical solution to the problem we established that when an electrostatically stabilized suspension is exposed to shear forces, it will first undergo a regime characterized by a lag time, the duration of which depends on the competition between repulsive electrostatic energy barrier between two particles and the shear-advection forces. In this regime, very limited aggregation is observed. However, the initially slow formation of clusters, which are much more sensitive to the presence of shear forces than primary particles, leads to a progressive acceleration of the aggregation kinetics (auto-catalytic regime). Such acceleration is highly non-linear, and typically culminates with an explosive, runaway behavior of the cluster growth. For sufficiently large clusters, the electrostatic repulsion becomes negligibly important, and they aggregate at the same shear-controlled rate as without repulsion. Experimental data support this picture, and indicate the existence of an additional lag time between the cluster-size explosion (as measured by light scattering) and the runaway of other measureable quantities, less sensitive to the presence of a few clusters, such as the suspension viscosity. Of great interest is that the combination of this peculiar aggregation mechanism with shear induced cluster breakage leads to the emergence of well-defined bimodal cluster mass distributions, with one population of primary particles and very small clusters and a second population of large clusters, whose size is defined by the dynamic balance between aggregation and breakage. We have discussed the most important features of the population balance equations, including the intrinsic stiffness of the resulting equations and how to circumvent it. The numerical predictions have been compared with an ample set of experimental data, from which it emerges that the developed model is able to account for the unique runaway behavior of viscosity of electrostatically-stabilized colloidal dispersions subject to shear forces, with good quantitative agreement. We believe that the proposed model could have a broad impact in clarifying previously unexplained phenomena, in particular clogging phenomena in microfluidics[50, 51] and contributes significantly in showing the rich behavior that the application of steady shear flow induces in ubiquitous colloidal systems such as those described by DLVO theory.

**Acknowledgments**

Financial support from the Swiss National Science Foundation (Grant number PP00P2133597/1) is gratefully acknowledged. A.Z. acknowledges financial support from the TUM Institute for Advanced Study, funded by the 191 German Excellence Initiative and the EU 7th Framework Programme under Grant 192 Agreement No. 291763.

**Figures**

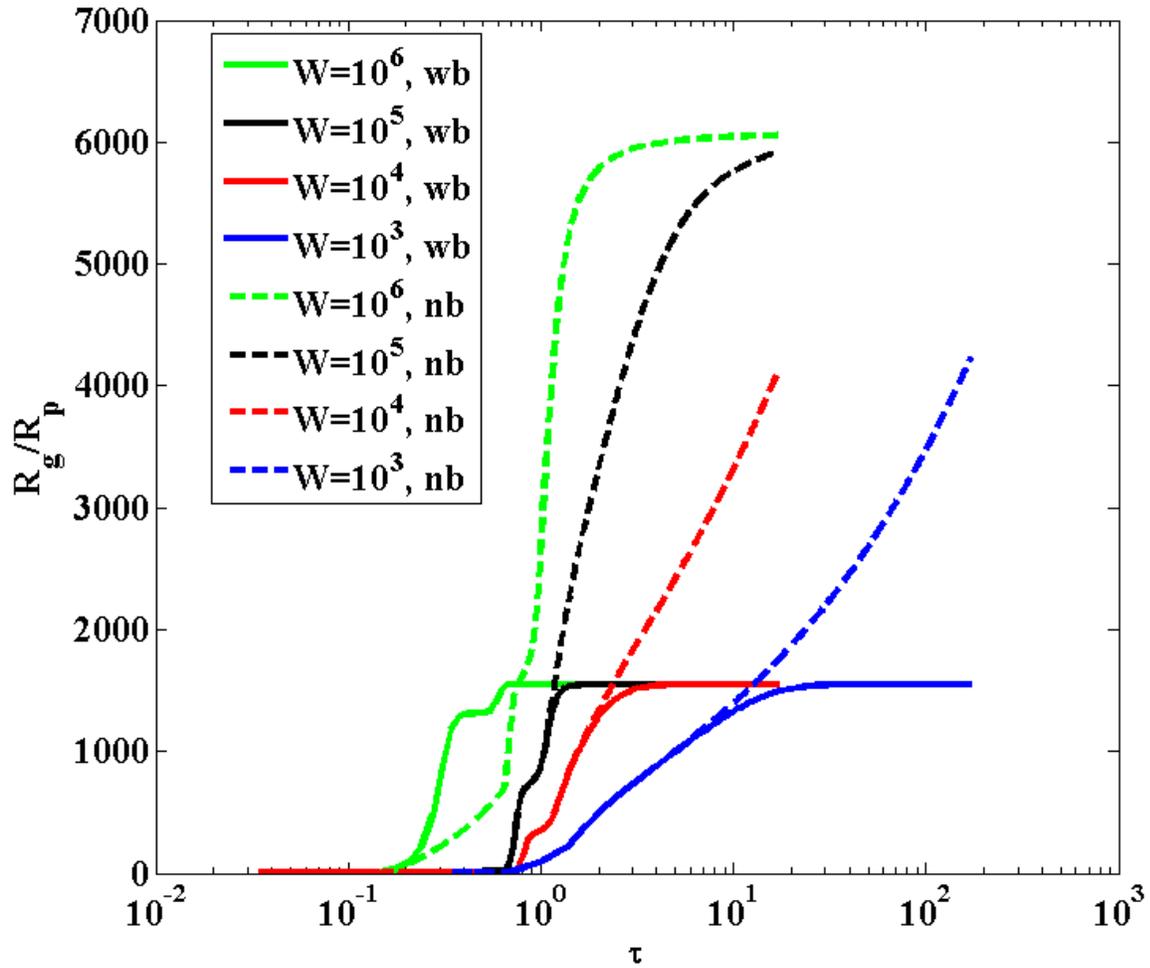

**Figure 1** Dimensionless radius of gyration evolution as a function of dimensionless time, for four different stability ratio values *W*, indicated in the legend, when the aggregation is modeled by Kernel (2). Both calculations with breakage (wb, modeled using Kernel (6)), and without breakage (nb) are reported. The calculations have been carried out for a particle diameter equal to 120 nm, particle volume fraction equal to 21% and shear rate equal to 1700 s$^{-1}$.

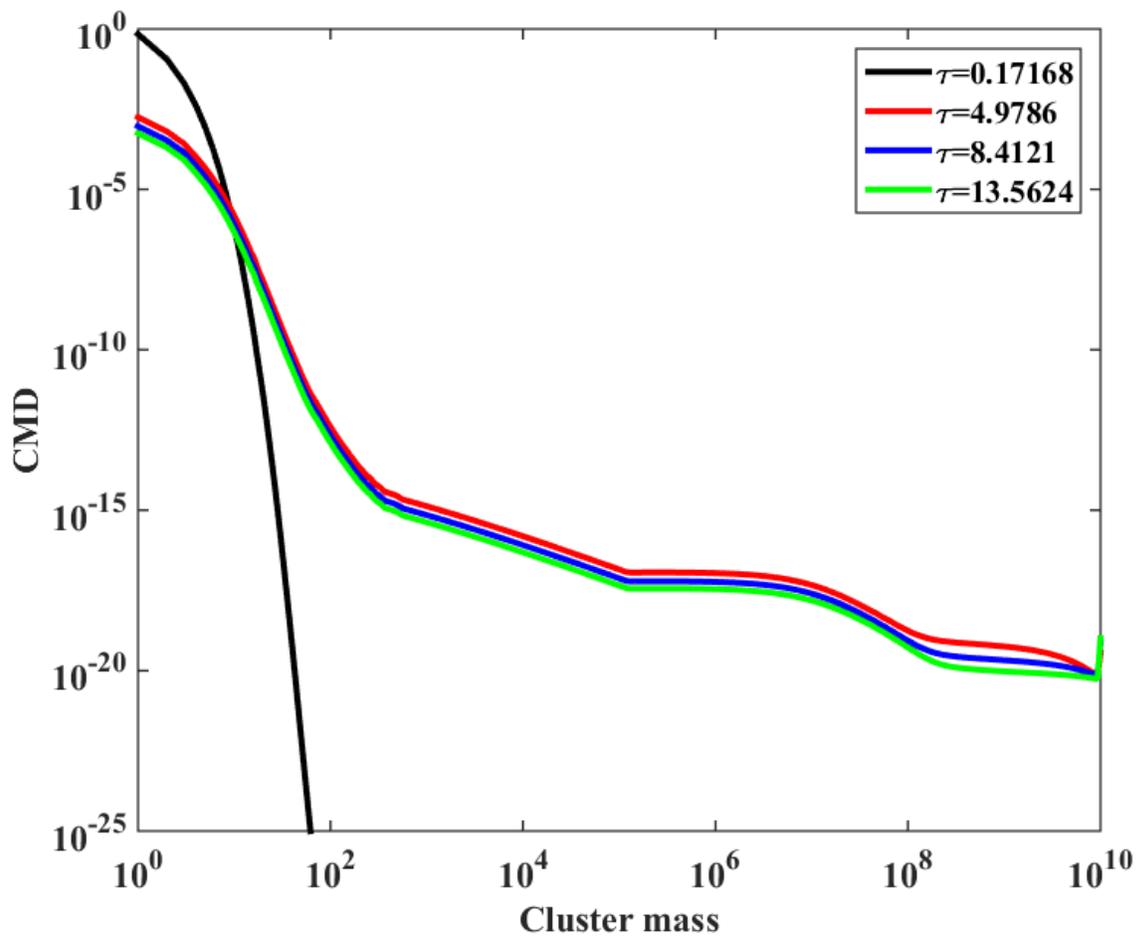

**Figure 2a** Cluster mass distribution as a function of the cluster mass (expressed as the number of primary particles), for four dimensionless times indicated in the legend, in the case where the aggregation is modeled by Kernel (2). The calculations have been carried out for particle diameter equal to 120 nm, $W=10^5$, particle volume fraction equal to 21% and shear rate equal to 1700 s$^{-1}$. The prefactor in the breakage rate constant in Equation (6) is equal to $c_1=0$ (i.e., no breakage).

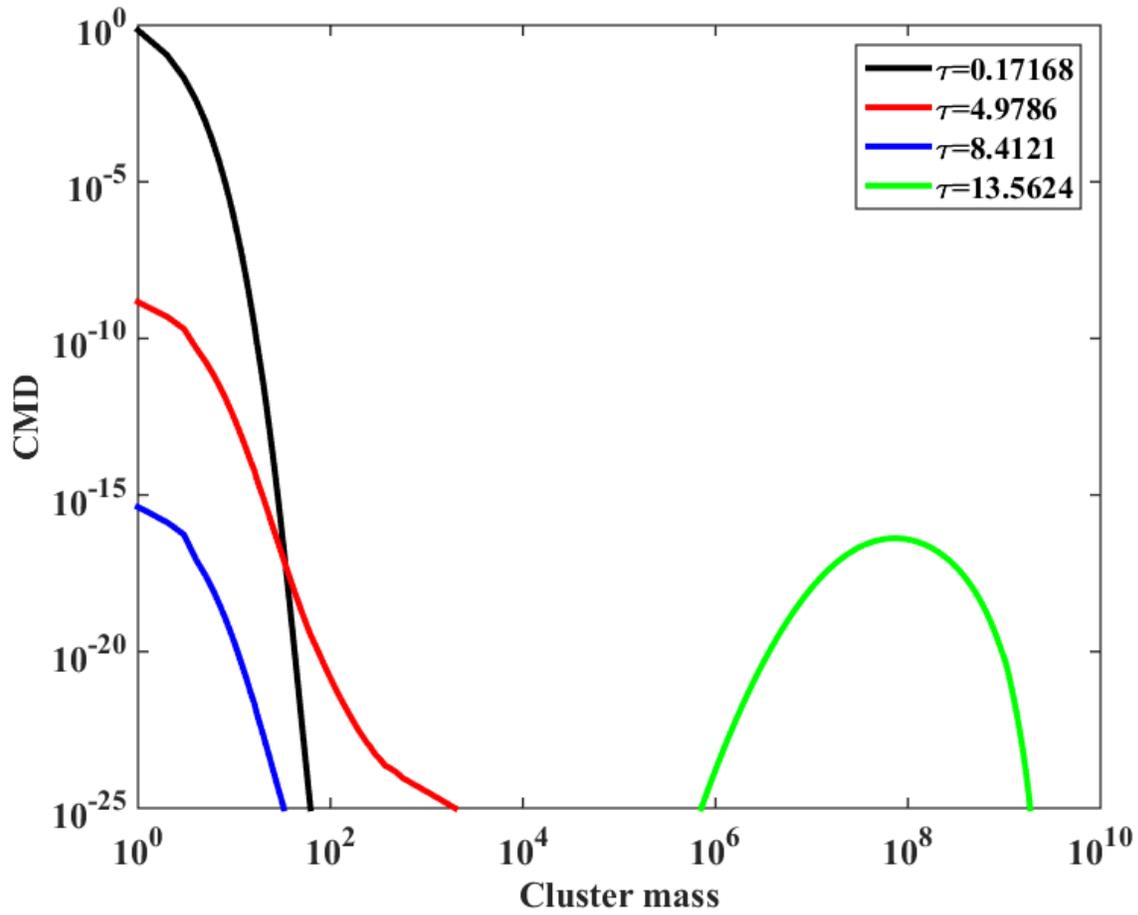

**Figure 2b** Cluster mass distribution as a function of the cluster mass (expressed as the number of primary particles), for four dimensionless times indicated in the legend, in the case where the aggregation is modeled by Kernel (2). The calculations have been carried out for particle diameter equal to 120 nm, $W=10^5$, particle volume fraction equal to 21% and shear rate equal to 1700 s$^{-1}$. The prefactor in the breakage rate constant in Equation (6) is equal to $c_1=2.38·10^{-10}$.

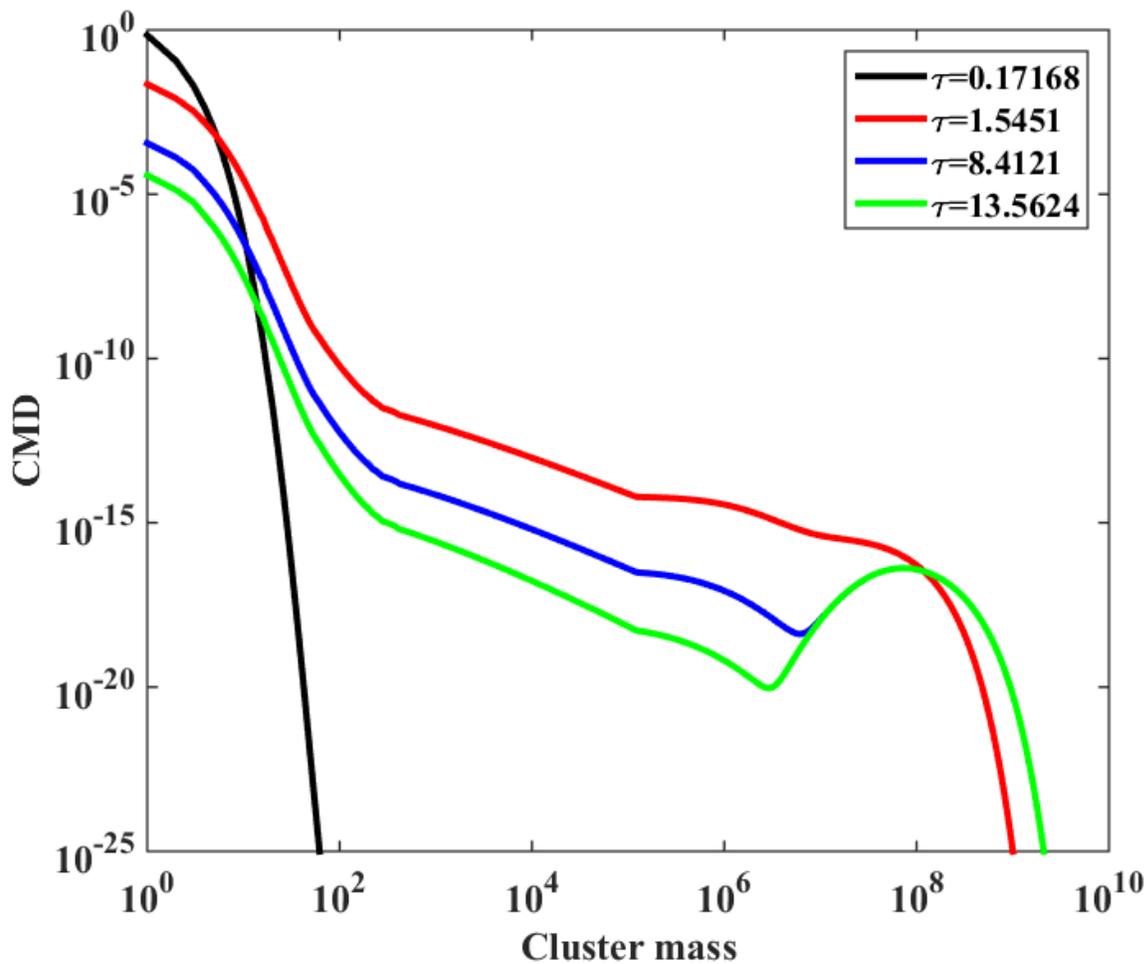

**Figure 3a** Cluster mass distribution as a function of the cluster mass (expressed as the number of primary particles), for four dimensionless times indicated in the legend, in the case where the aggregation is modeled by Kernel (2). The calculations have been carried out for particle diameter equal to 120 nm, $W=10^4$, particle volume fraction equal to 21% and shear rate equal to 1700 s$^{-1}$. The prefactor in the breakage rate constant in Equation (6) is equal to $c_1=2.38 \cdot 10^{-10}$.

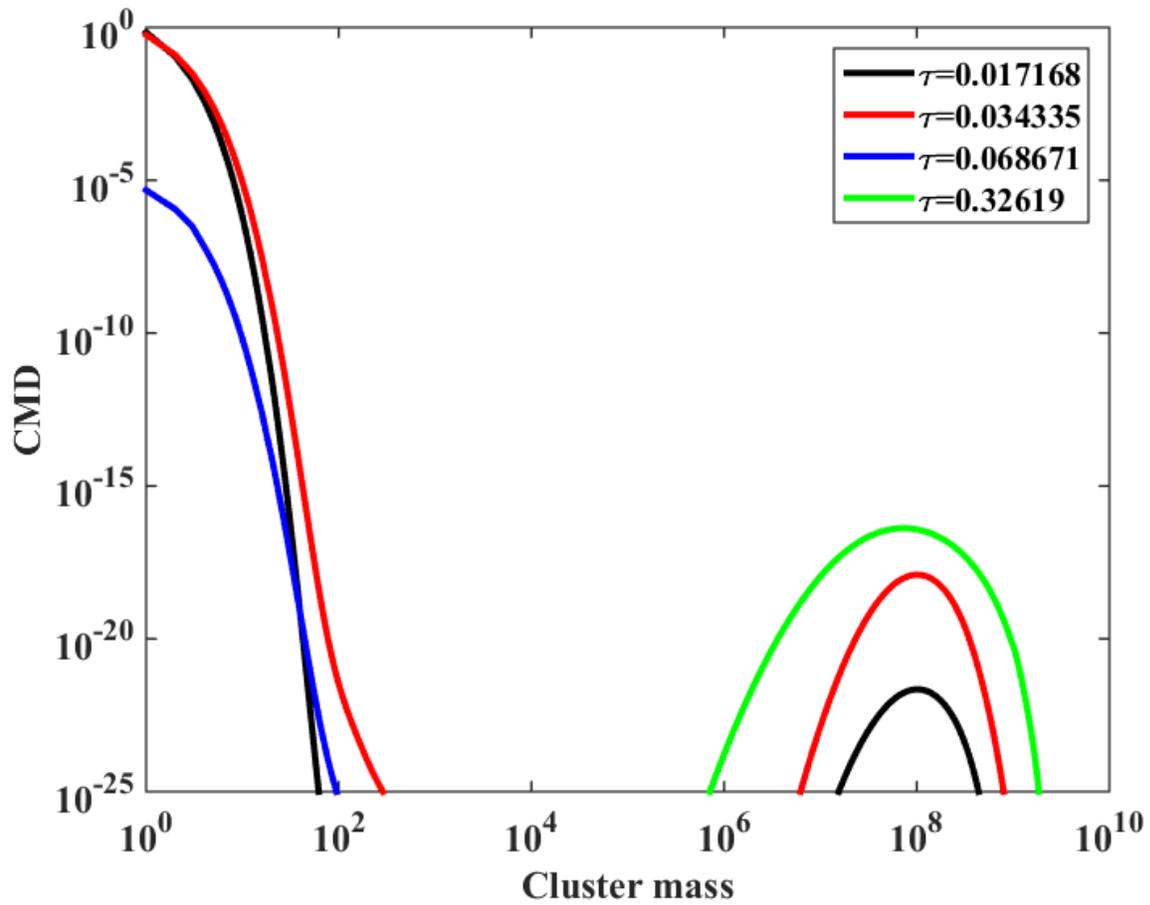

**Figure 3b** Cluster mass distribution as a function of the cluster mass (expressed as the number of primary particles), for four dimensionless times indicated in the legend, in the case where the aggregation is modeled by Kernel (2).. The calculations have been carried out for particle diameter equal to 120 nm, $W=10^6$, particle volume fraction equal to 21% and shear rate equal to 1700 s$^{-1}$. The prefactor in the breakage rate constant in Equation (6) is equal to $c_1=2.38 \cdot 10^{-10}$.

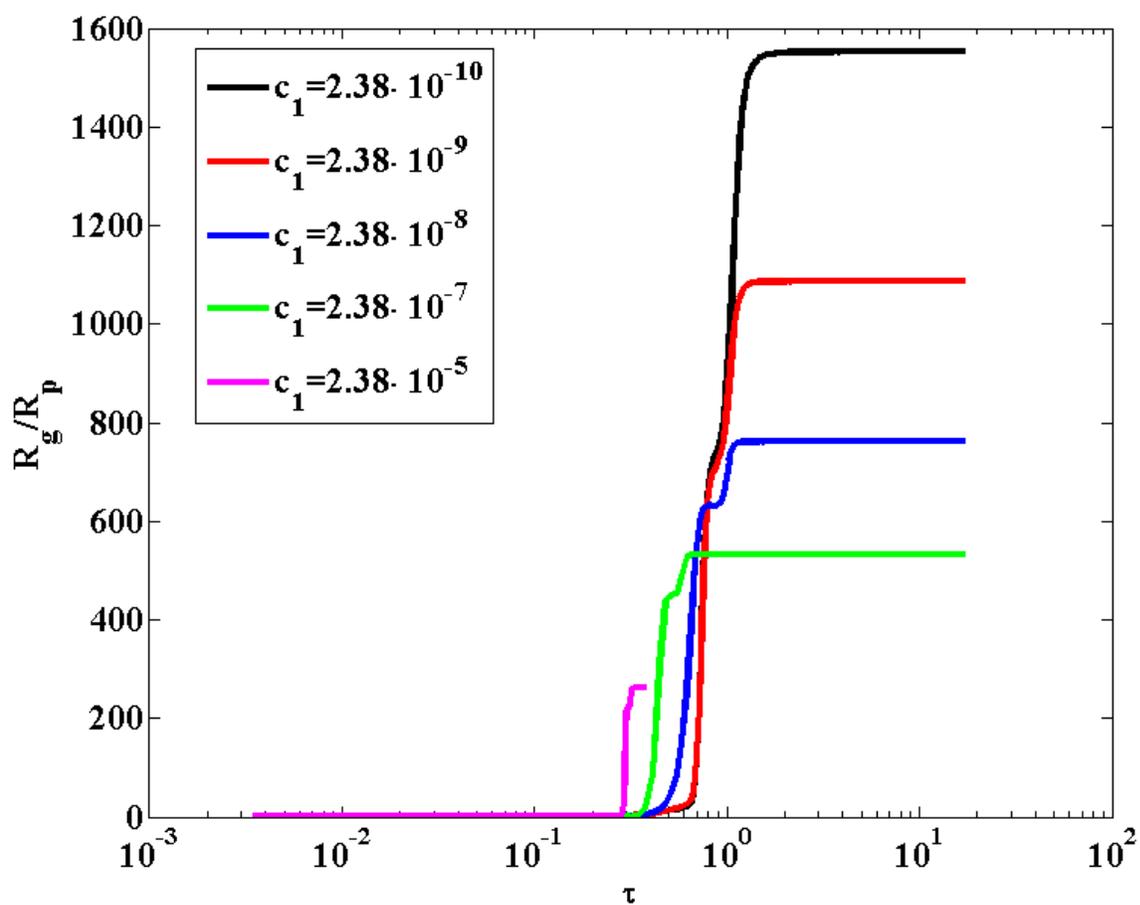

**Figure 4** Dimensionless radius of gyration evolution as a function of dimensionless time, in the case where the aggregation is modeled by Kernel (2). for five different values of the breakage rate constant prefactor $c_1$ in Equation (6) indicated in the legend. The calculations have been carried out for particle diameter equal to 120 nm, $W=10^5$, particle volume fraction equal to 21% and shear rate equal to 1700 s$^{-1}$.

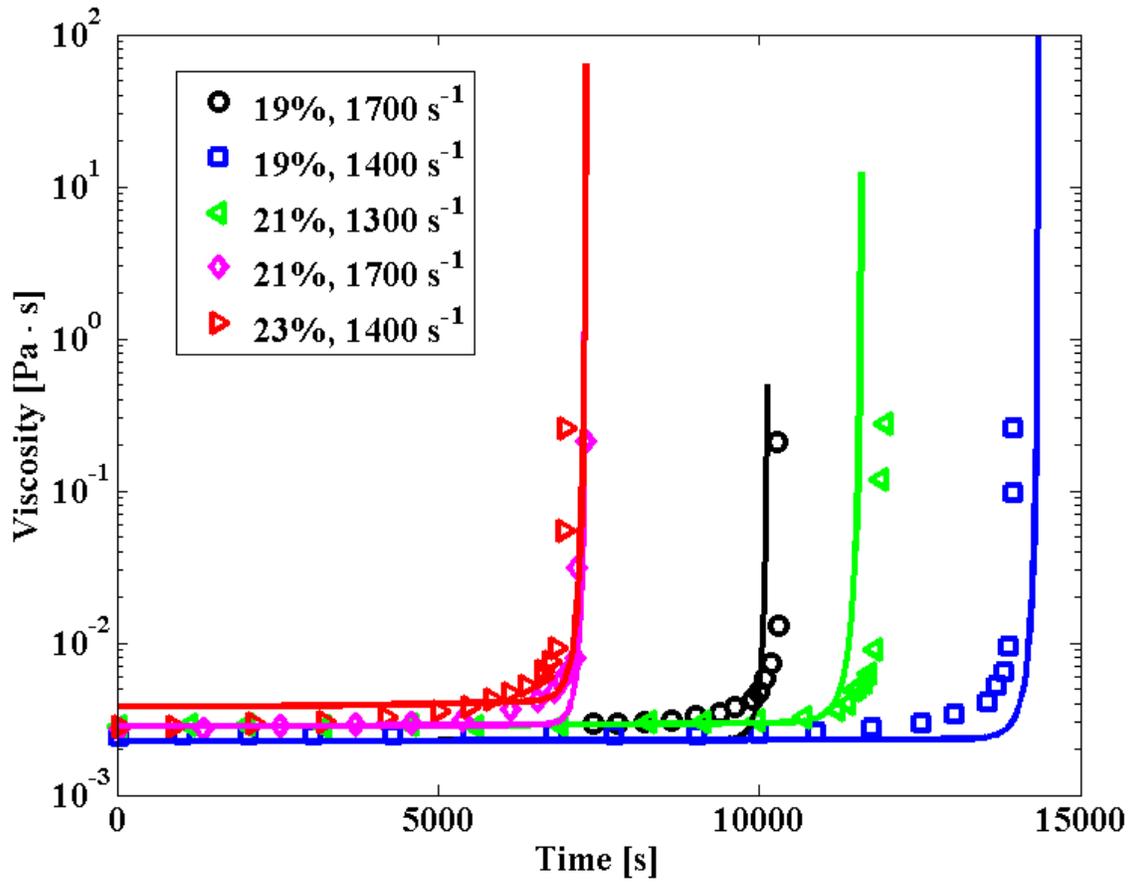

**Figure 5** Suspension viscosity evolution profiles as a function of time, for four different shear rates and particle volume fractions, as indicated in the legend. The points are experimental data, the lines the corresponding model predictions, in the case where the aggregation is modeled by Kernel (2). The calculations have been carried out with the following stability ratio values: *W=1.38·10$^8$* for particle volume fraction equal to 19%, *W=10$^8$* for particle volume fraction equal to 21% and *W=6.5·10$^7$* for particle volume fraction equal to 23%.

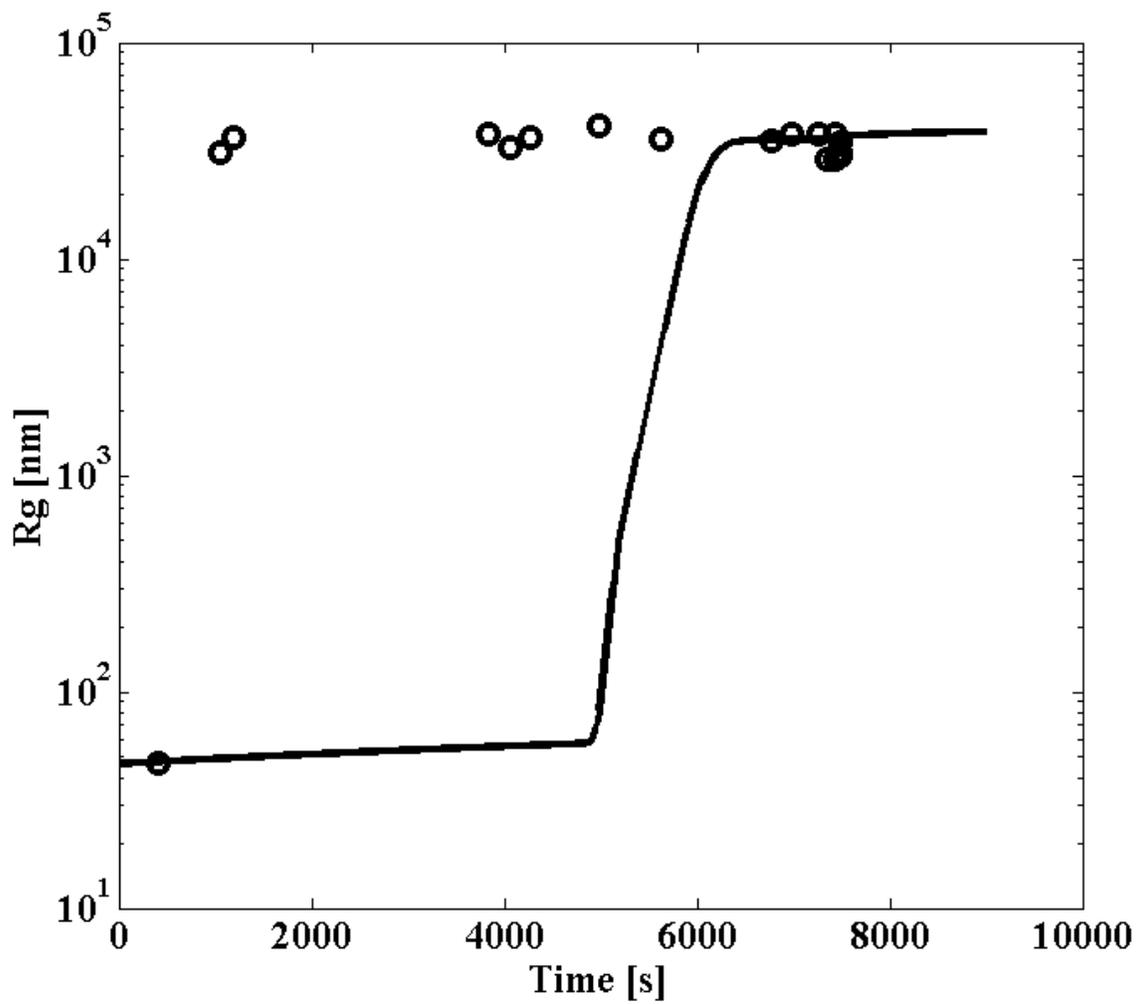

**Figure 6** Radius of gyration evolution as a function of time, for particle volume fraction equal to 21% and shear rate of 1700 s$^{-1}$. The points are experimental data, while the line is the corresponding model predictions, in the case where the aggregation is modeled by Kernel (2). The calculations have been carried out with the stability ratio value *W=10$^8$*.

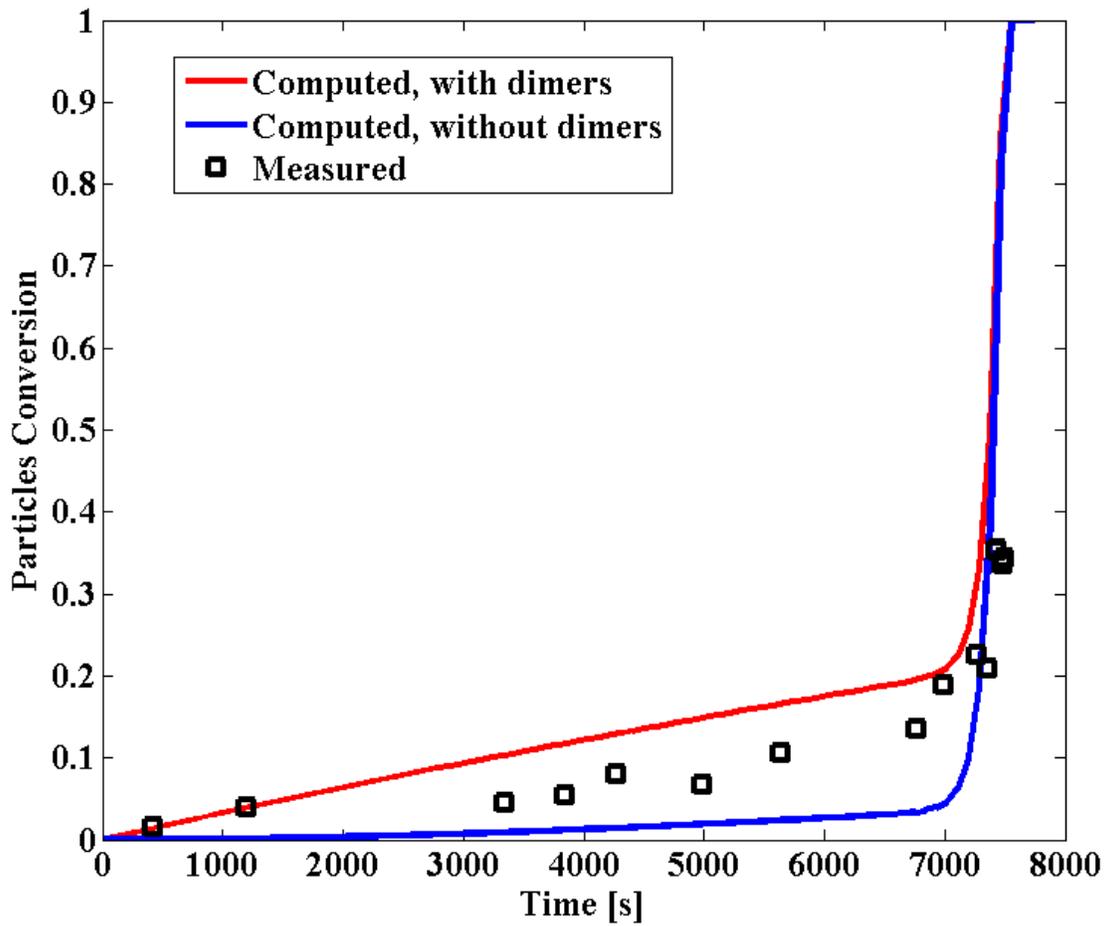

**Figure 7** Conversion of particles into large clusters for particle volume fraction equal to 21% and shear rate of 1700 s$^{-1}$. The points are experimental data, while the lines are the corresponding model predictions, in the case where the aggregation is modeled by Kernel (2). The red line corresponds to the conversion to all clusters, including dimers, while the blue line corresponds to the conversion to all clusters, excluding dimers. The calculations have been carried out with the stability ratio value *W=10$^8$*.

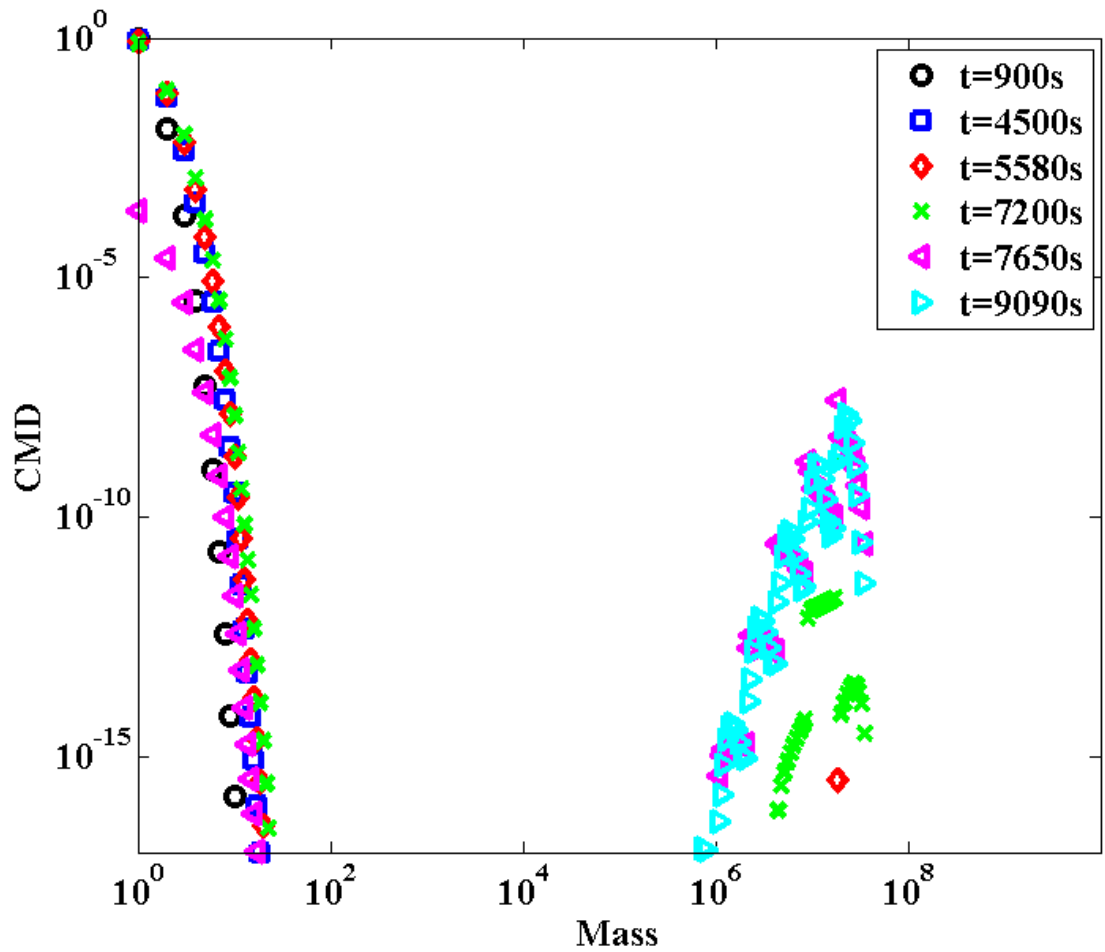

**Figure 8** Cluster mass distribution as a function of the cluster mass (expressed as the number of primary particles), for six times indicated in the legend, in the case where the aggregation is modeled by Kernel (2). The calculations have been carried out for particle diameter equal to 120 nm, $W=10^8$, particle volume fraction equal to 21% and shear rate equal to 1700 s$^{-1}$.

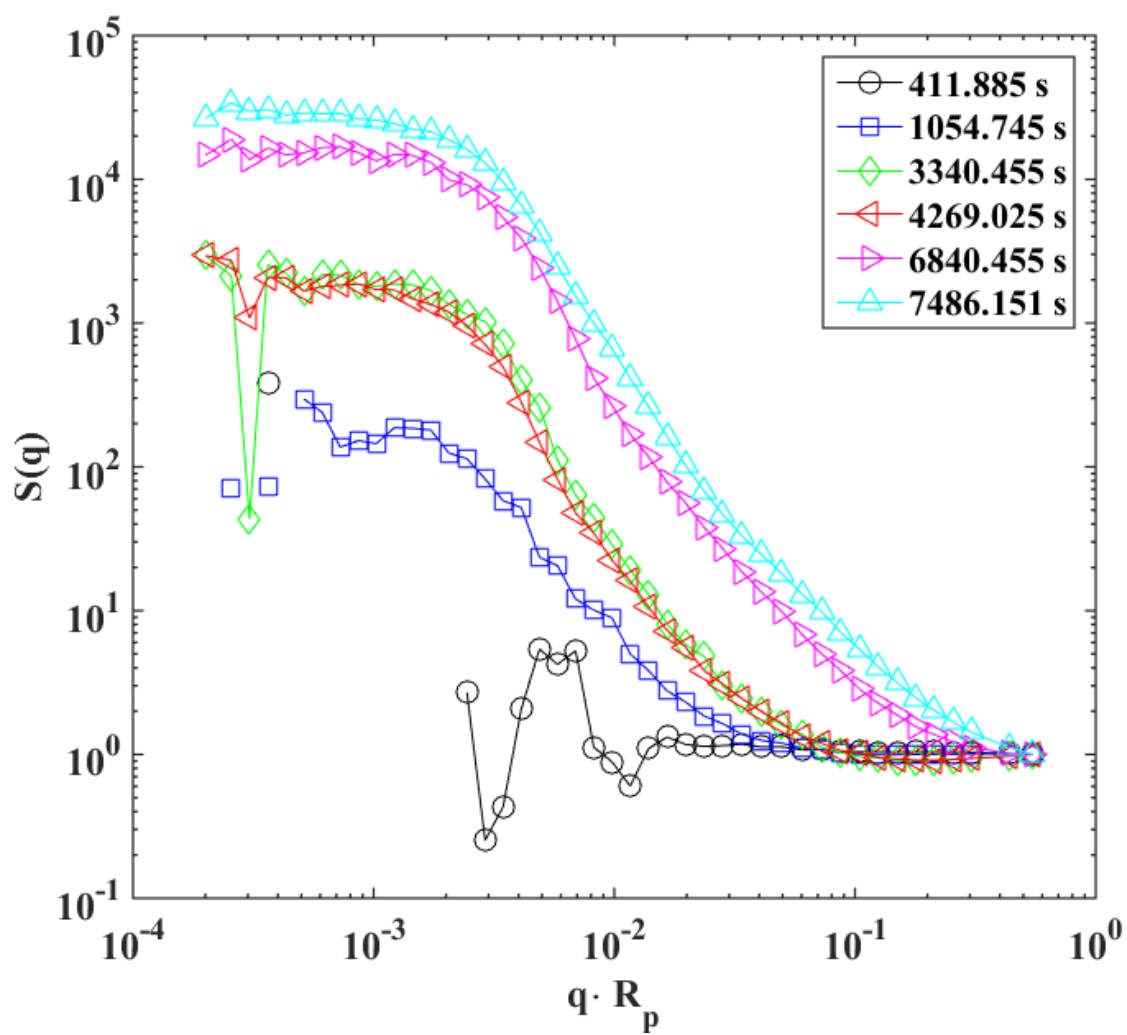

**Figure S9a** Experimental average Scattering Structure Factor as a function of the dimensionless scattering wave vector, for the times indicated in the legend. The data have been collected for a particle volume fraction equal to 21%, and a shear rat of 1700 s$^{-1}$.

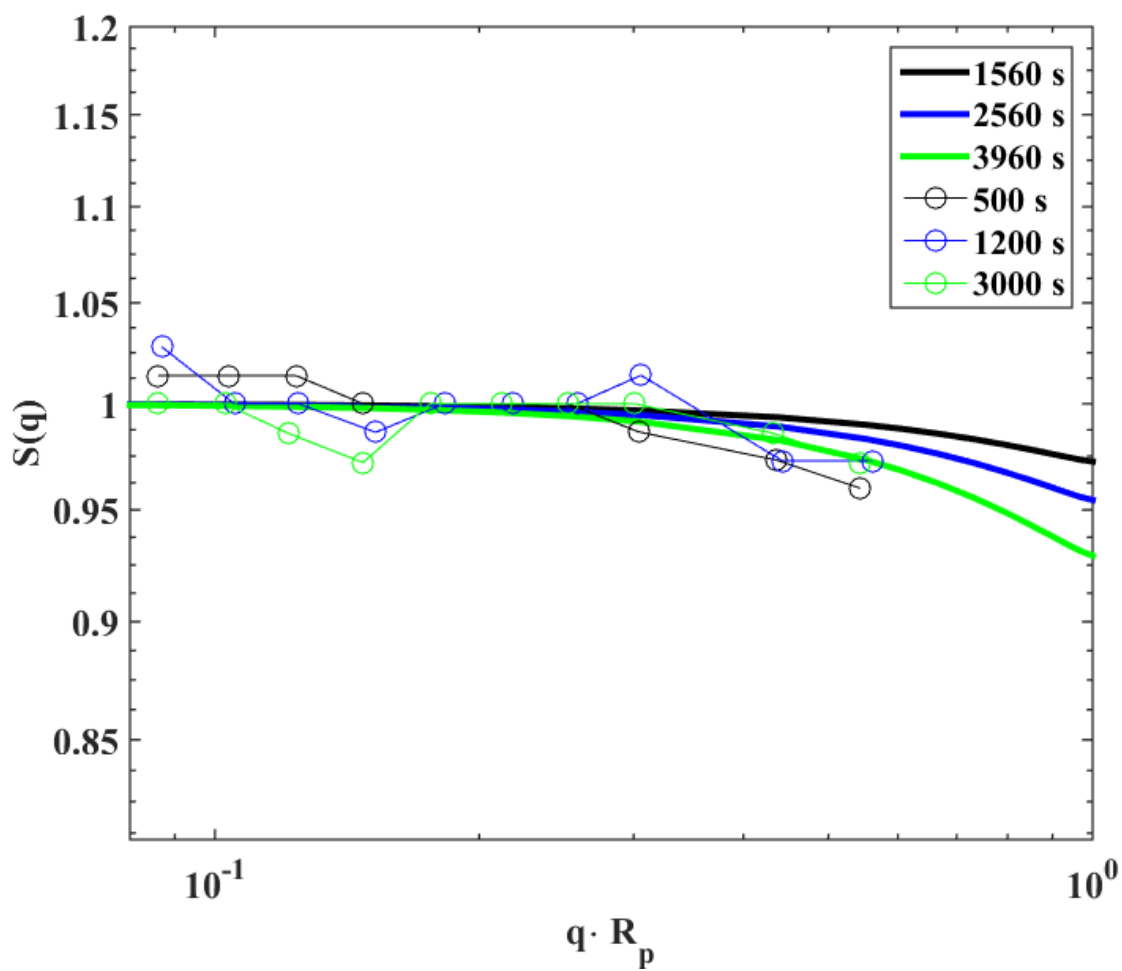

**Figure S9b** Experimental and calculated average Scattering Structure Factors as a function of the dimensionless scattering wave vector, for the times indicated in the legend. The experimental data have been obtained after filtering the suspension with a 5 micrometer filter, and the calculated structure factors have been computed by excluding all clusters with a diameter larger than 5 micrometers, in the case where the aggregation is modeled by Kernel (2). The data have been collected for a particle volume fraction equal to 21%, and a shear rat of 1700 s$^{-1}$, as well as $W=10^8$.

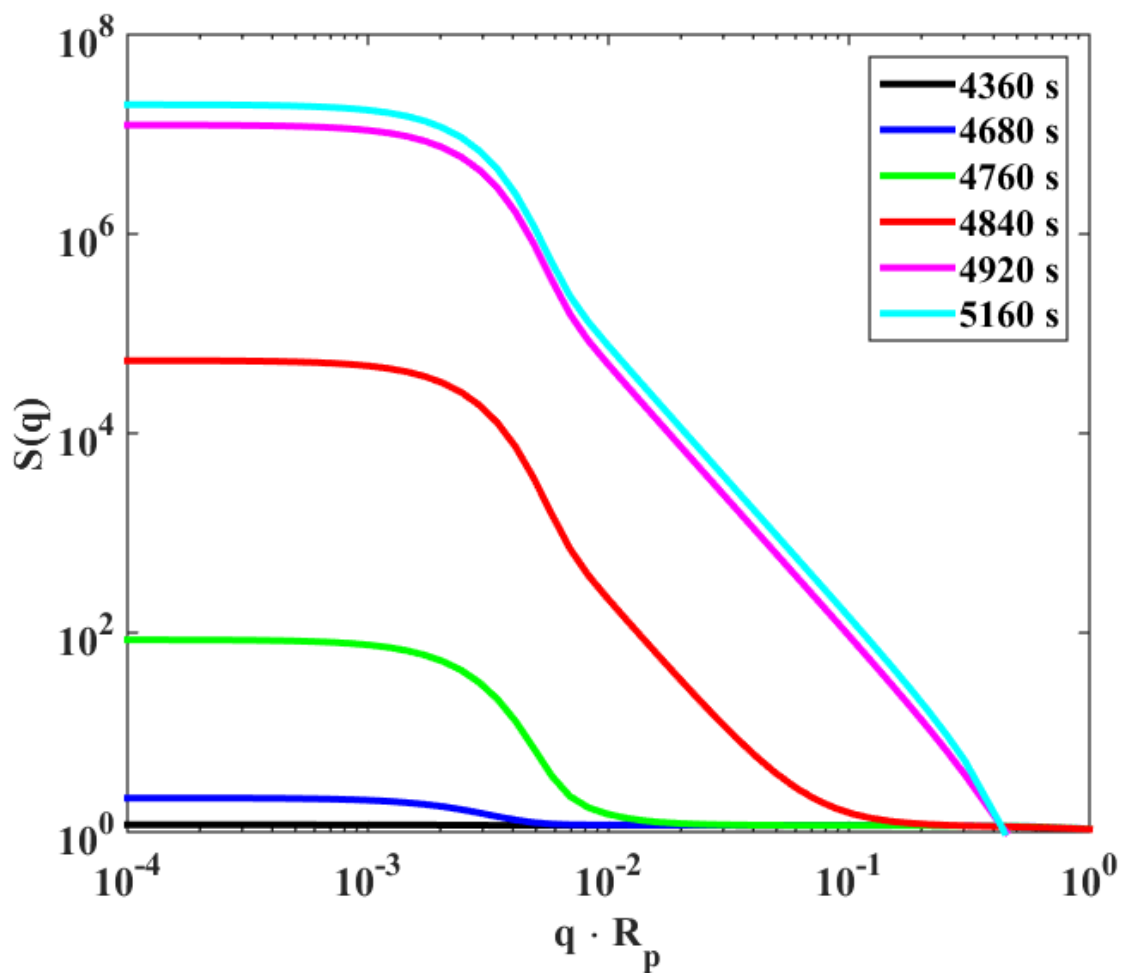

**Figure S9c** Calculated average Scattering Structure Factors as a function of the dimensionless scattering wave vector, for the times indicated in the legend, in the case where the aggregation is modeled by Kernel (2). The calculations have been carried out for a particle volume fraction equal to 21%, and a shear rat of 1700 s$^{-1}$, as well as $W=10^8$.

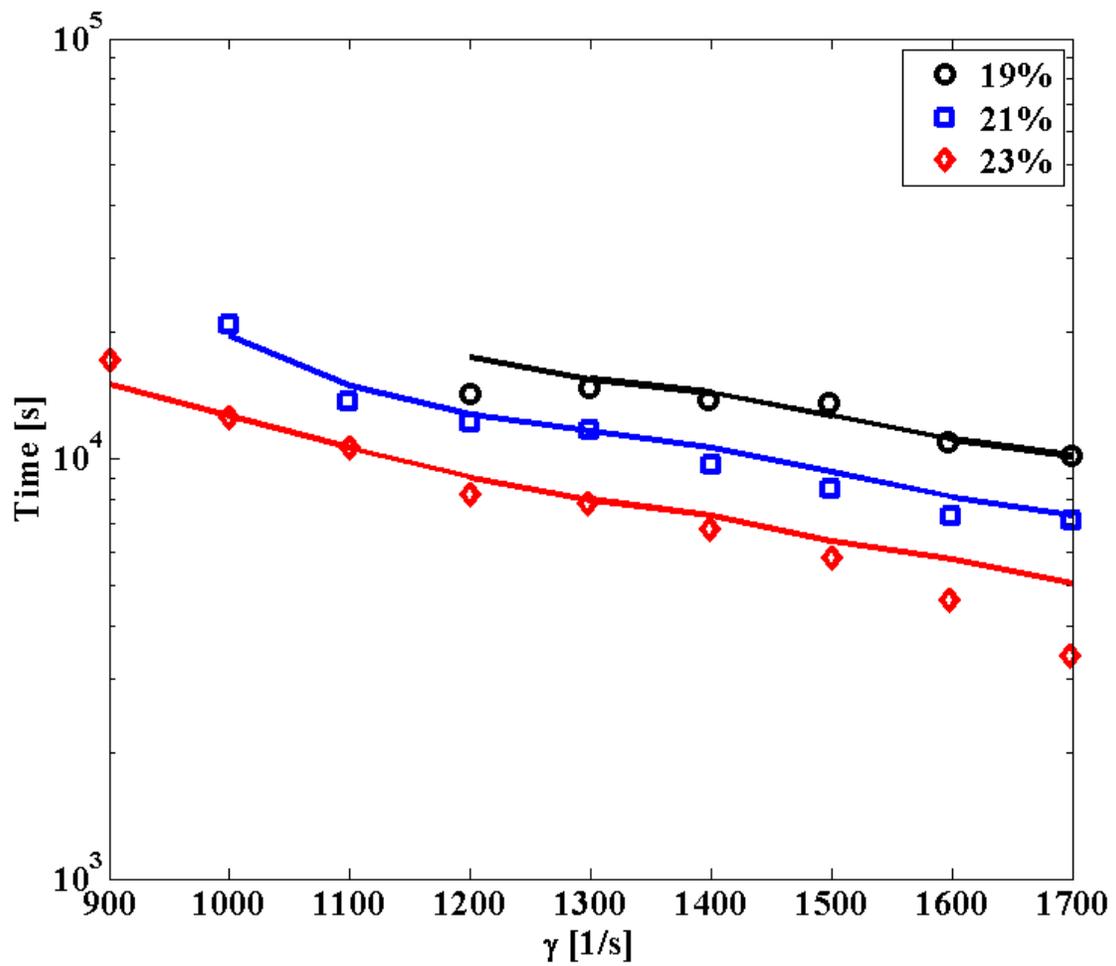

**Figure 10** Explosion times determined from the viscosity profiles as a function of the shear rate, for three different particle volume fractions, as indicated in the legend. The points are experimental data, while the lines are the corresponding model predictions. The calculations have been carried out with the following stability ratio values: $W=1.38 \cdot 10^8$ for particle volume fraction equal to 19%, $W=10^8$ for particle volume fraction equal to 21% and $W=6.5 \cdot 10^7$ for particle volume fraction equal to 23%.